\documentclass[%
 reprint,
superscriptaddress,
 amsmath,amssymb,
 aps,
pra,
]{revtex4-1}

\usepackage{hyperref}
\hypersetup{colorlinks=true, citecolor=blue, urlcolor=blue, linkcolor=blue}
\usepackage{graphicx}
\usepackage{dcolumn}
\usepackage{bm}
\usepackage{lipsum} 

\begin{document}


\title{Ab initio Determination of Phase Stabilities of Dynamically Disordered Solids: rotational C$_2$ disorder in Li$_2$C$_2$
}

\author{Johan Klarbring}
 \email{johan.klarbring@liu.se}
 \affiliation{%
Theoretical Physics Division, \\
Department of Physics, Chemistry and Biology (IFM),
Link\"{o}ping University, SE-581 83, Link\"{o}ping, Sweden
}%
\author{Stanislav Filippov}
 \affiliation{%
Department of Materials and Environmental Chemistry, \\ Stockholm University,106 91 Stockholm, Sweden
}%
\affiliation{%
Theoretical Physics Division, \\
Department of Physics, Chemistry and Biology (IFM),
Link\"{o}ping University, SE-581 83, Link\"{o}ping, Sweden
}%
\author{Ulrich H\"{a}ussermann}
 \affiliation{%
Department of Materials and Environmental Chemistry, \\ Stockholm University,106 91 Stockholm, Sweden
}%
\author{Sergei I. Simak}
\email{sersi78@liu.se}
 \affiliation{%
Theoretical Physics Division, \\
Department of Physics, Chemistry and Biology (IFM),
Link\"{o}ping University, SE-581 83, Link\"{o}ping, Sweden
}%
\affiliation{%
Department of Physics and Astronomy,
\\
Uppsala University, SE-75120 Uppsala, Sweden
}%

\date{\today}

\begin{abstract}
The temperature-induced orthorhombic to cubic phase transition in Li$_2$C$_2$ is a prototypical example of a solid to solid phase transformation between an ordered phase, which is well described within the phonon theory, and a dynamically disordered phase with rotating molecules, for which the standard phonon theory is not applicable. The transformation in Li$_2$C$_2$ happens from a phase with directionally ordered C$_2$ dimers to a structure, where they are dynamically disorderd. We provide a description of this transition by employing ab initio molecular dynamics (AIMD) based stress-strain thermodynamic integration on a deformation path that connects the ordered and dynamically disordered phases. The free energy difference between the two phases is obtained.  The entropy that stabilizes the dynamically disordered cubic phase is captured by the behavior of the stress on the deformation path. We also show that a combination of the stress-strain thermodynamic integration and machine learning force field methodologies appears as a promising path in studies of dynamically disordered materials.

\end{abstract}

\pacs{Valid PACS appear here}
\maketitle

\section{\label{sec:Intro} Introduction}
The ability to accurately predict and understand solid-solid phase transformations is a longstanding goal in theoretical solid state physics. 
Predicting and ultimately controlling the external conditions (temperature (T), pressure (P) or applied fields) at which these phase transformations occur is of great importance for a host of applications. Indeed, the transformations may be detrimental to performance, or may be the phenomenon on which a device could be based, as eg.\ in mechanocaloric materials for solid-state cooling \cite{Cazorla2019} and shape-memory alloys \cite{MohdJani2014}.

In order to predict the critical temperature of the phase transformations, accurate determination of the relevant free energies is required. First-principles electronic structure methods based on density functional theory (DFT) have been very successful in predicting pressure induced phase transformations at low temperatures, where the free energy is well approximated by the enthalpy. 

At relatively low T (as compared to melting) many solids are well described within the harmonic approximation. That is, the atomic vibrations in the system are well represented by a set of non-interacting harmonically oscillating normal-modes, and accurate predictions of temperature-induced structural transitions can thus be made, see eg. Ref. \cite{Fultz2010}. This approach is particularly accurate if extended to the so-called quasi-harmonic approximation, where harmonic phonons are calculated at thermally expanded volumes.

In many cases, however, the (quasi-)harmonic approximation is insufficient. For instance for systems that are unstable with respect to the equilibrium atomic positions of some high symmetry phase at 0~K, but where these positions are stabilized by anharmonic phonon-phonon interactions at elevated temperatures. In such anharmonic cases, re-normalized phonon schemes, including self-consistent phonons \cite{Souvatzis2008,Erra2014,Bianco2017,Tadano2015}, and effective harmonic models fitted from molecular dynamics (MD) \cite{Hellman2011,Hellman2013}, have proven to be effective. 

Corrections to the free energy of an (effective) harmonic model from higher order terms in the phonon expansion can also be derived \cite{wallace1998} and have been recently applied in a first-principles context \cite{Ravichandran2018}.

Beyond these "phonon-based" techniques, the concept of thermodynamic integration (TI) provides a set of formally exact methods to extract free energies, which are increasingly being used in connection with first-principles DFT-related techniques \cite{Grabowski2009,Duff2015,Wilson2013}. Methods in this class are based on the fact that while the full free energy, or more precisely the entropy, is very difficult to extract directly from an MD simulation, certain derivatives of the free energy can be more easily extracted and the free energy itself can then be extracted by integration over several different simulations. This is founded on the basic result that free energy derivatives can be evaluated as thermal averages over energy derivatives, i.e.\ for an external parameter $\lambda$ coupled to the system it holds that \cite{Landau_stat_book}
\begin{equation}
\label{eq:TI_basis}
    \left( \frac{\partial F(V,T,\lambda)}{\partial \lambda} \right) = \overline{ \frac{\partial H(\{\mathbf{q},\mathbf{p}\};\lambda)}{\partial \lambda}},
\end{equation}
where H is he Hamiltonian of the sytem. The phase-space ($\{\mathbf{q},\mathbf{p}\}$) ensemble average on the right-hand-side is practically evaluated as a time-average over an MD simulation under the assumption of ergodicity. 

The most commonly used TI method is the so called Kirkwood coupling constant integration. Here, one adds to the free energy a parametrical dependence on an external parameter $\lambda$, i.e., $F = F(V,T,\lambda)$ and the free energy change upon changing $\lambda$ is obtained by integration of the right-hand-side of Eq.\ (\ref{eq:TI_basis}).

A less commonly employed TI technique involves obtaining free energy changes by integrating pressure over a volume change, or more generally, as will be the basis of the calculations in this work, integrating a stress related quantity over a deformation path connecting two (real or artificial) phases of interest \cite{Haskins2016,Klarbring2018,Ozolins2009}.  

A certain set of solids, which we will refer to as dynamically disordered solids, necessitates the use of free energy methods beyond phonon based ones. These are systems which possess some type of time-averaged long-range order, characteristic of a crystalline solid, but where well-defined, unique, equilibrium positions cannot be statically assigned to each individual atom.
\begin{figure*}[ht!]
    \includegraphics[trim={1cm 0.5cm 0cm 0cm},clip,width=1.5\columnwidth]{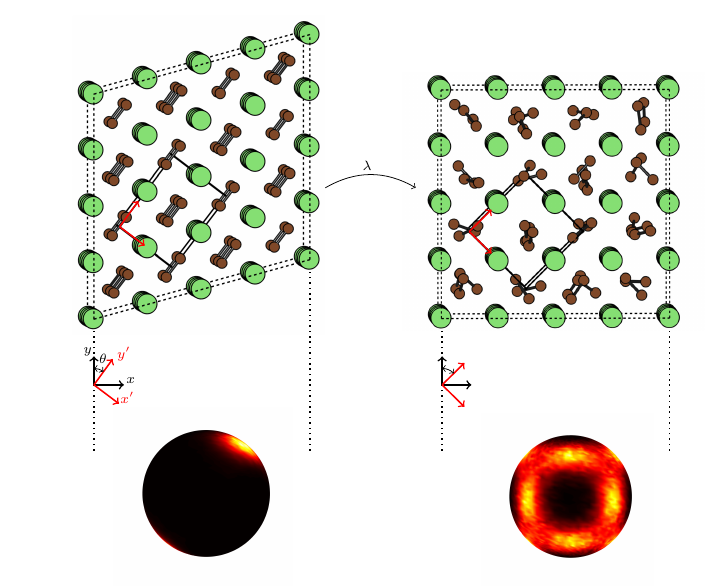}
    \caption{\label{fig:struct} Illustration of the deformation path (parametrized by $\lambda$) from the ordered orthorhombic (left) to the dynamically disordered cubic (right) phase with the supercells used in this work. Li and C atoms are colored green and brown, respectively. In the cubic phase a snapshot of the positions of C atoms are taken from an AIMD simulation, in order to illustrate the rotational disorder of the C$_2$ dimers, while the Li atoms are displayed in their high-symmetry positions. The conventional orthorhombic unit cell is illustrated by solid black lines. The Cartesian reference coordinate system, $\{\mathbf{x},\mathbf{y},\mathbf{z}\}$, used in our simulations is indicated by black vectors and letters, while the coordinate system, $\{\mathbf{x'},\mathbf{y'},\mathbf{z'}\}$, aligned with the conventional orthorhombic cell is displayed in red. $\mathbf{z}$ and $\mathbf{z}'$ are collinear and directed out of the figure. The spheres at the bottom schematically represent the angular distribution of the C$_2$ dimers over AIMD simulations at 600 K. Bright colors and dark colors correspond to high density and low density, respectively.}
\end{figure*}

They involve certain fast ion-conducting solids, so called superionics, which posses a "liquid-like" diffusional behavior of certain ionic species \cite{Hull2004}. Such systems are being increasingly intensively studied, with envisioned applications ranging from solid electrolytes in batteries and fuel cells to thermoelectric materials \cite{Wang2015,Liu2012}.

Another clear example of dynamically disordered solids are those containing rotating molecular units, sometimes referred to as plastic or orientation-disordered solids. Such systems typically have a high temperature phase where the center of mass of each molecular unit is ordered on certain high symmetry positions, but where the individual atoms making up the molecule are positionally disordered. To cite an emerging application, we mention that some such systems are highly promising barocaloric materials for solid-state cooling applications \cite{Li2019_baro,Cazorla2019,Lloveras2019,Cazorla2019_2}.

In this work, we investigate from first principles the temperature induced orthorhombic-to-cubic phase transformation of Li$_2$C$_2$ at ambient pressure. We choose Li$_2$C$_2$ as it is a very "clean" example of a system with a temperature-induced phase-transformation between an ordered and a dynamically disordered phase with rotating molecules. The low temperature orthorhombic phase of Li$_2$C$_2$ is ordered with C$_2$ dimers aligned along the orthorhombic $b$ axis. Upon heating, this phase transforms into the cubic phase, where the C$_2$ dimers are dynamically disordered among shallow local minima corresponding to alignment along the [110] directions of the cubic structure \cite{Filippov2019}. Fig.\ \ref{fig:struct} shows the orthorhombic and cubic phases and the nature of the dimer alignment in these phases. We employ the stress-strain TI framework to investigate the nature of this transformation. We find that the stabilizing entropy associated with the onset of disorder among the C$_2$ dimers is seen through the behavior of the stresses on the deformation path between the two phases, and fair agreement with experiments for the critical temperature of the transition.

\section{Methodology}
\subsection{Li$_2$C$_2$ Polymorphs and Stress-Strain Thermodynamic Integration}
Figure \ref{fig:struct} illustrates the supercells used to represent the low-temperature orthorhombic and high temperature cubic phases. The relation between the monoclinic and the conventional orthorhombic representation of the low-temperature phase is displayed. Viewed in the coordinate system $\{ \mathbf{x}',\mathbf{y}',\mathbf{z}' \}$, aligned with the orthorhombic axes, the transformation is a compression along $\mathbf{y}'$ and an expansion along $\mathbf{x}'$ and $\mathbf{z}'$. The supercell matrices used for the two phases in this work are given in the Appendix.

The stress-strain TI methodology starts from the differential of the Helmholtz free energy, $F$, for a rotationally invariant crystal \cite{Wallace1970,Klarbring2018},
    \begin{equation}
\label{eq:diff_F}
 dF(\{\mathbf{X},\boldsymbol{\eta}\},T) = -SdT + V(\mathbf{X}) \boldsymbol{\tau} \mathbin{\mathbf{:}} d\boldsymbol{\eta} .
\end{equation}
where $S$ and $T$ are the entropy and temperature, $\mathbf{X}$ is an arbitrary reference configuration, $\boldsymbol{\eta}$ is the Lagrangian strain tensor, defined w.r.t $\mathbf{X}$, and $V(\mathbf{X})$ and $\boldsymbol{\tau}$ are the volume and the second Piola-Kirchoff stress tensor, respectively.

Integrating along a deformation path parametrized by $\lambda$ at constant T finally yields the free energy change as \cite{Klarbring2018}
\begin{equation}
\label{eq:DeltaF_general}
   \Delta F(\lambda) = \int_0^\lambda V(\lambda') \boldsymbol{\sigma}(\lambda') \boldsymbol{h}^{-T}(\lambda')  \mathbin{\mathbf{:}} \frac{\partial \boldsymbol{h}(\lambda')}{\partial \lambda'} d \lambda',
\end{equation}
where $\boldsymbol{\sigma}$ is the Cauchy stress tensor, related to $\boldsymbol{\tau}$ as $\boldsymbol{\tau} = \mathbin{J}  \boldsymbol{\alpha}^{-1} \boldsymbol{\sigma} \boldsymbol{\alpha}^{-T}$. The deformation gradient $\boldsymbol{\alpha}$ is the derivative of each component of a material point in the
configuration $\boldsymbol{x}$ with respect to each component of the material point in the reference configuration $\boldsymbol{X}$, and $ J = \det(\boldsymbol{\alpha})$. For $\boldsymbol{x}$ = $\boldsymbol{x}(\boldsymbol{X})$, the elements of $\boldsymbol{\alpha}$ are

\begin{equation}
\alpha_{ij}  =  \frac{\partial x_i}{\partial X_j}.  
\end{equation}
We define the matrix $\boldsymbol{h}$ to contain as its columns the supercell lattice vectors $\boldsymbol{a}$, $\boldsymbol{b}$ and $\boldsymbol{c}$ used in the AIMD simulations,
\begin{equation}
    \boldsymbol{h} =     
    \begin{pmatrix}
     a_x & b_x & c_x \\
     a_y & b_y & c_y \\
     a_z & b_z & c_z \\
    \end{pmatrix}.
\end{equation}
For homogeneous deformations, which is what we treat here, $\boldsymbol{\alpha}$  relates the lattice vectors in the initial configuration $\boldsymbol{X}$ and a configuration $\boldsymbol{x}$ as
\begin{equation}
\label{eq:deform_h}
    \boldsymbol{h}(\boldsymbol{x}) = \boldsymbol{\alpha} \boldsymbol{h}(\boldsymbol{X}).
\end{equation}

$\boldsymbol{\eta}$ is related to $\boldsymbol{\alpha}$ as 
\begin{equation}
\label{eq:deff_strain}
 \boldsymbol{\eta} = \frac{1}{2} \left( \boldsymbol{\alpha}^T\boldsymbol{\alpha} - \boldsymbol{I} \right), 
\end{equation}
where  $\boldsymbol{I}$ is the identity matrix.

We find it revealing to analyse the Cauchy stress in the coordinate system aligned with the conventional orthorhombic cell, $\{ \mathbf{x}',\mathbf{y}',\mathbf{z}' \}$, see Fig.\ \ref{fig:struct}. The Cauchy stress $\boldsymbol{\sigma}$ transforms as $\boldsymbol{\sigma}' = \mathbf{Q}\boldsymbol{\sigma}\mathbf{Q}^{T}$, where $\mathbf{Q}$ is the transformation matrix from the coordinate system $\{\mathbf{x},\mathbf{y},\mathbf{z}\}$ to $\{ \mathbf{x}',\mathbf{y}',\mathbf{z}' \}$ given by 
\begin{equation}
    \mathbf{Q} = 
\begin{pmatrix}
\cos\theta & -\sin\theta & 0\\
\sin\theta & \cos\theta & 0 \\
0 & 0 & 1
\end{pmatrix},
\end{equation}
where $\theta$ is defined in Fig.\ \ref{fig:struct}. Note that $\theta$ and hence $\mathbf{Q}$ changes along the deformation path, i.e., they are $\lambda$ dependent. 

\subsection{Computational details}
We carried out Kohn-Sham (KS) DFT calculations using the projector augmented wave (PAW) \cite{blochl1994} method and a plane wave basis set as implemented in the Vienna ab initio simulation package (VASP) \cite{kresse1996,kresse1996_2,kresse1999}. Exchange and correlation effects were treated using the Perdew-Burke-Ernzerhof (PBE) \cite{Perdew1996} form of the generalized gradient approximation (GGA). PAW potentials with Li(1s2s) and C(2s2p) states treated as valence were used. The plane-wave cutoff energy of the KS orbitals was 600 eV. We used a 128 atom supercell constructed as a 2$\times$2$\times$2 expansion of the conventional cubic anti-fluorite unit cell, for which k-point sampling was performed at the $\Gamma$-point. This k-point sampling is enough to converge the stress tensor within $\sim$ 1 kbar, as judged by comparing to $2 \times 2 \times 2 $ and $3 \times 3 \times 3$ k-point grids on 10 snapshots from AIMD simulations. AIMD simulations in the NVT ensemble were performed using a short 0.25 fs timestep and a Nos\'{e}-Hoover thermostat with a Nos\'{e} mass corresponding to a time-constant of $\sim$ 86 timesteps. The presence of $C_2$ dimer stretching modes at around 60 THz, separated from the rest of the vibrational modes in the system,  makes it necessary to carefully choose the timestep and the Nos\'{e} mass in order to avoid having the C and Li subsystems slowly decoupled and equilibrate at two separate temperatures over long timescales. This was observed to happen using more standard parameters, e.g.\ a 1 fs timestep and the default VASP value for the Nos\'{e} mass. The combined simulation time at each point on the deformation path was typically at least $\sim$360000 timesteps, corresponding to 90 ps of simulation time, and for select cases significantly longer to judge convergence. We typically skip at least $\sim$ 40 ps as equilibration. More rationale on the simulation times follows below.

To allow for long simulation times we employ a Gaussian approximation potential (GAP) \cite{Bartok2010} machine learned force field (MLFF) in the form implemented in VASP, including the associated on-the-fly learning procedure, and use this to cross-check certain AIMD results for convergence with respect to simulation time. Full technical details of the implementation are described in Ref.\ \cite{Jinnouchi2020} 

We largely use the VASP  default values (as of version 6.3.0) of the hyperparameters. For instance, a 5 Å cutoff for both the radial and angular descriptors used to represent local atomic environments and a 0.5 Å width of the Gaussian broadening of the atomic distributions.

The on-the-fly learning MD runs are performed with a 0.5 fs timestep in the same 128 atom supercell, as follows. We first heat the system from 300 to 800 K in a variable cell MD setting over 200000 timesteps, corresponding to a heating rate of 5 K/ps. During this heating run the system transforms from the orthorhombic to the cubic phase at around ~665 K. We then continue the learning in the NpT ensemble for 100000 timesteps at 800 K in the cubic phase. Finally, we perform 100000 timesteps of NVT MD at each of 3 equally spaced points on the deformation path connecting the orthorhombic and cubic phase at 600 K. From these MD runs a total of 1136 DFT snapshots were added to the training set, from which 708 and 1667 local reference configurations for Li and C, respectively, were selected.  

Finally, we refit the MLFF by solving the associated linear regression problem using singular value decomposition (SVD) and, since we are primarily concerned with stresses, increasing the weight of the stresses in the fitting problem by 10 times compared to the energies and forces.  

The thus obtained MLFF yields mean absolute errors (MAE) on the training set of 0.87 meV/atom, 28 meV/\AA\ and 0.17 kbar for the energies, force components and stress tensor components. We also construct a test set of 200 snapshots, 100 from the cubic and 100 from the othorhombic phases at 600 K. The MAEs on this test set are 0.52 meV/atom, 26 meV/\AA\ and 0.28 kbar for the energies, force components and stress tensor components, respectively.

\section{Results and discussion}

\subsection{Thermal expansion}
\begin{figure}[ht!]
    \includegraphics[trim={0cm 0cm 0.55cm 0.35cm},clip,width=\columnwidth]{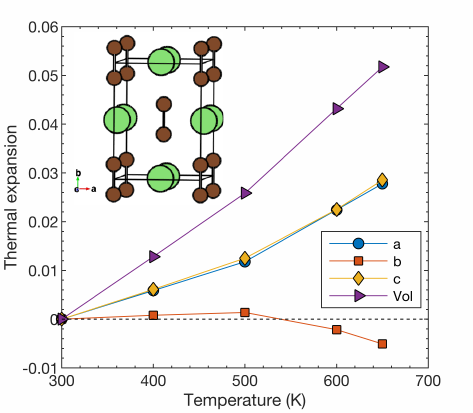}
    \caption{\label{fig:thermal_exp} Thermal expansion of the $a$, $b$ and $c$ lattice parameters and volume of the orthorhombic structure of Li$_2$C$_2$. The thermal expansion is evaluated as $(a(T)-a(300 \text{ K}))/a(300 \text{ K})$ for $a$ and similar for $b$, $c$, and the volume. Inset shows the conventional orthorhombic unit cell.}
\end{figure}
We start by investigating the temperature dependence of the lattice constants in the orthorhombic and cubic phases of Li$_{2}$C$_{2}$. We extract these from AIMD by iteratively changing the lattice constants until all elements of $\boldsymbol{\sigma}$ are $<$2 kBar (more details on this are discussed below). In practice this is done by taking the average stress from an AIMD run in the monoclinic cell, transforming it to the orthorhombic cell and changing the $a$, $b$, and $c$ lattice vectors according to the diagonal components of this transformed stress tensor in a steepest decent fashion. The procedure is then repeated until convergence.  

The obtained values at 300 K are $a =$ 3.68 \AA, $b =$ 4.83 \AA, and $c =$ 5.44 \AA, which agree very well with the experimental values at room temperature \cite{Ruschewitz1999}. Fig.\ \ref{fig:thermal_exp} shows the thermal expansion from T = 300 K, i.e., $(a(T)-a(300 \text{ K}))/a(300 \text{ K})$ and similar for $b$, $c$ and the volume V. While the $a$ and $c$ axes expand with temperature, the $b$ axis, which is collinear with the alignment of the $C_2$ dimers, shows a non-conventional decrease at high temperature. This is an effect of the temperature induced weakening of the alignment of the $C_2$ dimers. This peculiarity was pointed out experimentally for Li$_2$C$_2$ \cite{Ruschewitz1999}. 

\subsection{Stress-strain Thermodynamic integration}
\begin{figure}[ht!]
    \includegraphics[trim={0cm 0.3cm 0cm 0cm},clip,width=\columnwidth]{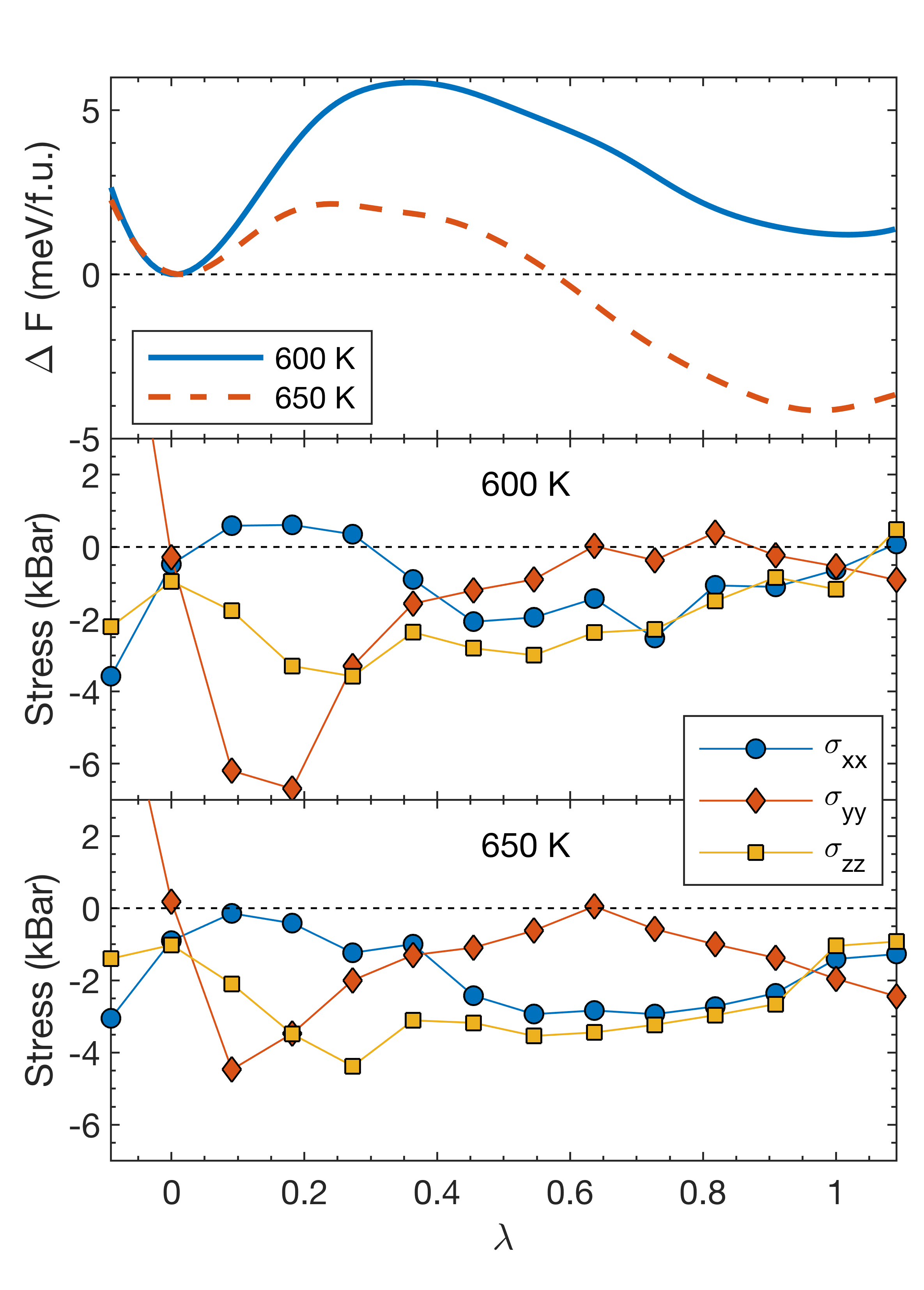}
    \caption{\label{fig:SS_TI} Results of the stress-strain thermodynamic integration on the deformation path from the ordered orthorhombic phase at $\lambda$ = 0 to the dynamically disordered cubic phase at $\lambda$ = 1. The top panel shows the free energy change along the deformation path at 600 K (solid blue line) and 650 K (dashed red line). Middle and lower panel show the diagonal components of the Cauchy stress in the coordinate system $\{ \mathbf{x}',\mathbf{y}',\mathbf{z}' \}$ aligned with the conventional orthorhombic unit cell, at 600 and 650 K, respectively.}
\end{figure}
We obtain the free energy difference $\Delta F(\lambda)$ on a deformation path between the ordered orthorhombic phase ($\lambda = 0$) and the disordered cubic phase ($\lambda = 1 $) using Eq.\ (\ref{eq:DeltaF_general}). We use 10 points obtained by linear interpolation between $\mathbf{h}(\lambda = 0)$ and $\mathbf{h}(\lambda = 1) $. Since the path connects an ordered phase with a dynamically disordered one, the disorder is expected to set in somewhere along the path. The convergence of the stress with respect to simulation time thus becomes a critical issue. The approach we take to attempt to ensure proper convergence is the following: At each temperature and each point on the deformation path we perform two separate AIMD runs, one with initial (fractional) atomic positions and velocities taken from an AIMD snapshot of the orthorhombic phase, and one with a snapshot from the cubic phase. Convergence can then be gauged by how well these two sets of stress-strain TIs agree with each other, as a function of simulation time. Indeed, insufficient simulation time in the two sets should produce errors of opposite type. Roughly speaking, the simulations started from the orthorhombic snapshot should be biased towards producing errors associated with too much order, while those started from the cubic snapshot should be biased towards too much disorder. We perform simulations long enough so that the integrand in Eq.\ (\ref{eq:DeltaF_general}) differs between the two AIMD runs by less than $\sim$ 1.6 meV/f.u at each point on the deformation path. The mean difference over the whole path is significantly smaller than this, around $\sim$ 0.5 meV/f.u.. The stresses used to produce the final results are averaged over the two simulations, in order to increase the accuracy further.

Fig.\ \ref{fig:SS_TI} (a) shows the free energy differences between the two phases at 600  and 650 K. We see that at 600 K the orthorhombic phase is favored by  1 meV/f.u., while at 650 K the cubic phase is 4 meV/f.u.\ lower in free energy. A linear interpolation results in a phase transition temperature of 611 K. Comparing this temperature to an experimental transition temperature is not straightforward. It was noted \cite{Ruschewitz1999} that the first appearance of an XRD reflection corresponding to the high temperature phase upon heating appears at around 693 K, and differential scanning calorimetry (DSC) measurements \cite{Dre2013} puts the transition at $\sim$ 725 K. Our thermodynamic transition temperature is thus likely underestimated, although in fair agreement considering all approximations inherent to the simulations, including the exchange-correlation functional and the finite and defect-free supercells. We note that this also highlights a well known difficulty in modelling solid-solid phase transitions, namely that small shifts in the free energy difference between the phases,  can lead to substantial shifts in the predicted transition temperature. As is clear from Fig.\ \ref{fig:SS_TI}, we clearly deal with small free energy differences, and the methodology is thus certainly susceptible to this difficulty.

It has also noted \cite{Ruschewitz1999}, however, that the two phases continue to co-exist over a broad temperature range. This is in agreement with the result from Fig.\ \ref{fig:SS_TI} (a) that the two phases appear to be free energy local minima both above and below the transition temperature.  

\begin{figure}[ht!]
    \includegraphics[trim={0cm 0.3cm 0cm 0cm},clip,width=\columnwidth]{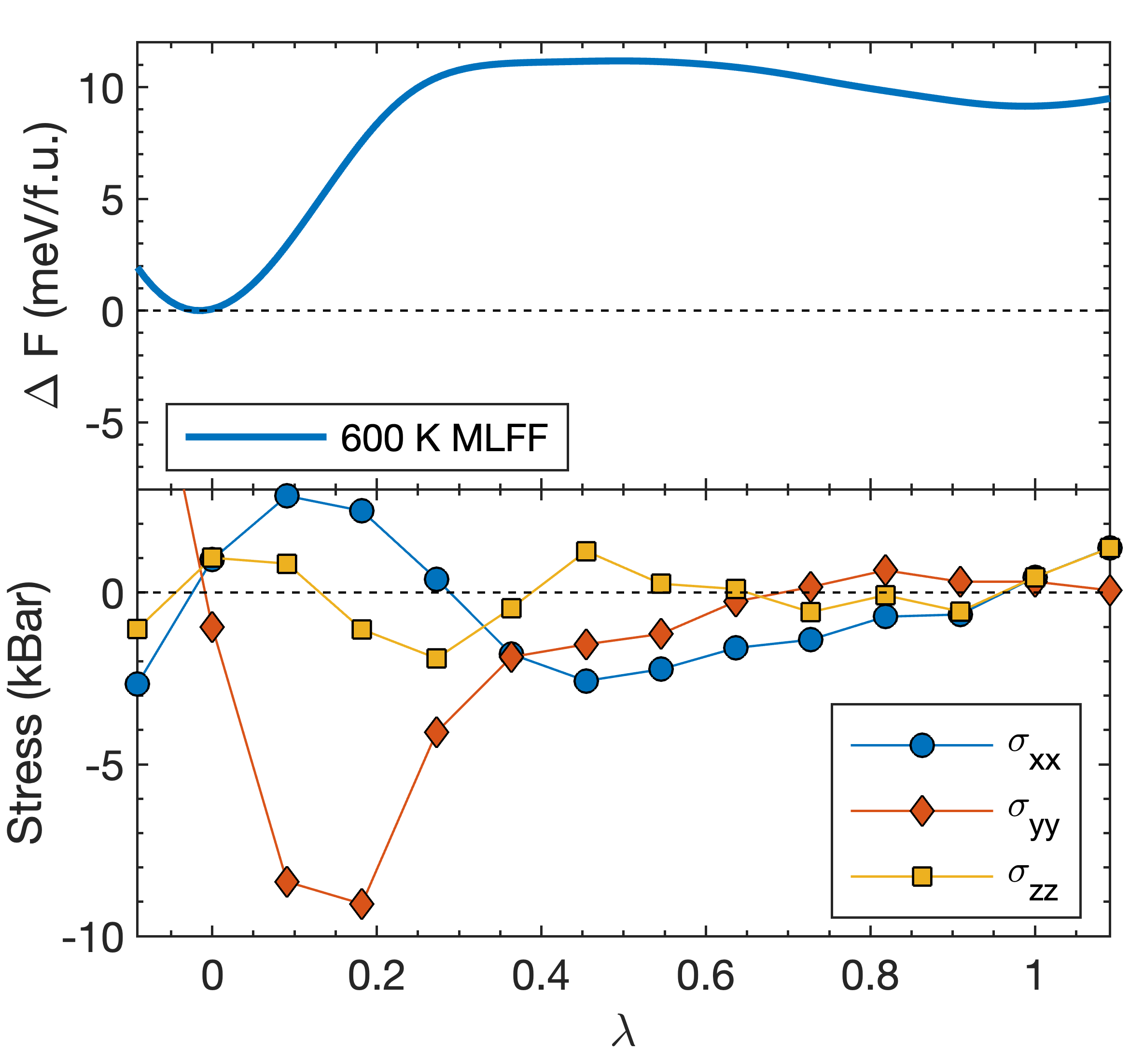}
    \caption{\label{fig:SS_TI_mlff} Results of the stress-strain thermodynamic integration at 600 K using the MLFF. The top panel shows the free energy change along the deformation path at 600 K (solid blue line). Lower panel show the diagonal components of the Cauchy stress in the coordinate system $\{ \mathbf{x}',\mathbf{y}',\mathbf{z}' \}$ aligned with the conventional orthorhombic unit cell, at 600, respectively.}
\end{figure}

Generally, the employed TI procedure may require large simulation times to properly capture the transition from order to dynamical disorder along the deformation path. Thus, machine-learning force-fields (MLFF) approaches, where the computational complexity of force evaluations can be reduced by orders-of-magnitude, may be of great help if they are able to provide the required accuracy. Though in this paper we do not rely on MLFFs for our main results, we have trained a GAP potential in the form implemented in VASP (see Section IIB), and use it to gauge if  qualitatively similar results as in the AIMD simulations are obtained using the MLFF with much larger simulations lengths. Fig. \ref{fig:SS_TI_mlff} shows the results of the SSTI at 600K using the same cell parameters as in the AIMD based run. We run each simulation for at least 2$\times$10$^6$ timesteps (500 ps), skipping 5$\times$10$^5$ (125 ps) for equilibration.  

We see that the qualitative behaviour of the stresses on the deformation path is the same as in the AIMD case (Fig.\ \ref{fig:SS_TI}). There are, however, some quantitative differences in the stresses, resulting in an over-stabilization of the orthorhombic phase, in comparison to the AIMD case. Nevertheless, we note that the qualitative picture of two minima, one for the ordered and one for the dynamically disordered phase, is reproduced by the MLFF, and thus holds in the limit of simulations times significantly longer than those accessible by AIMD.

The slight non-zero values of the stresses at $\lambda$ = 0 and 1 are due to the fact that the configurations chosen as starting points, turn out, once very long simulations have been run, to be slightly offset from the equilibrium ones. That is, the path we choose through configuration space passes slightly to the side of the actual minima. The change in free-energy in adjusting for this is, however, negligible, in the "worst" case, which is $\lambda$ = 1 at 650 K, this free energy correction can be estimated to be below 0.5 meV/f.u.

In order to gauge how the C$_2$ dimer disorder sets in on the deformation path we calculate a dimer self-correlation function,
\begin{equation}
    \label{eq:autcorr}
    C_{self}(t) =  \langle \tilde{\mathbf{S}}_i(t_0) \cdot \tilde{\mathbf{S}}_i(t+t_0) \rangle_{N,t_0},
\end{equation}
where $\tilde{\mathbf{S}}_i(t) = \mathbf{S}_i(t) - \langle\mathbf{S}_i(t)\rangle_N$, $\mathbf{S}_i(t)$ is the vector between the two C atoms making up C$_2$ dimer $i$ and $\langle \cdots \rangle_{N,t_0}$ denotes an average over both the $N$ C$_2$ dimers and all time-origins $t_0$.

\begin{figure}[ht!]
    \includegraphics[trim={0cm 0cm 0cm 0cm},clip,width=\columnwidth]{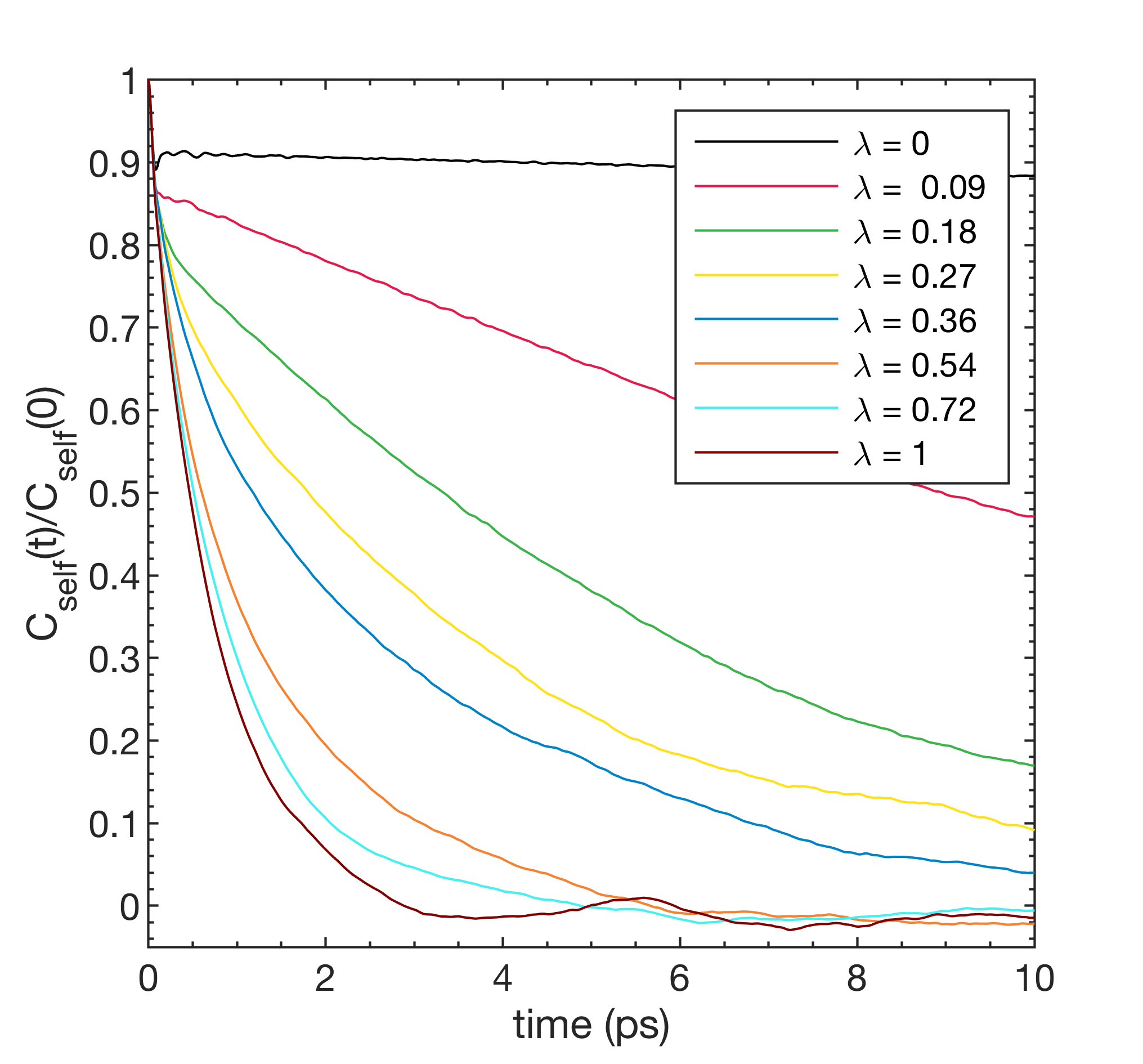}
    \caption{\label{fig:autocorr} Self-correlation function $C_{self}(t)/C_{self}(0)$ for a select number of points along the deformation path between the orthorhombic and cubic phases at 600 K.}
\end{figure}

In Fig.\ \ref{fig:autocorr} we plot $C_{self}(t)/C_{self}(0)$ for select values of $\lambda$ along the deformation path at 600 K. In the orthorhombic phase at $\lambda$= 0 we observe, as expected, a quick initial decay in the very short time limit followed by a virtually constant behaviour, consistent with the ordered state of the dimers. At the first point along the path ($\lambda$= 0.09) a similar initial decay is followed by a slow decay, indicating a still largely ordered state of the C$_2$ dimers. Moving further along the deformation path, the behaviour qualitatively changes into significantly faster decay rates. Around this point on the deformation path, we see a change in the behaviour of the $\sigma_{yy}$ stress component from the standard decrease upon compression, to an increase, which we can thus associate with the onset of proper disorder of the C$_2$ dimers, which effectively weakens the alignment of the dimers along the $y$ direction.

Starting from $\lambda$ = 0.27, $C_{self}(t)/C_{self}(0)$ is reasonably well fitted to a stretched exponential $e^{-(t/\tau)^\beta}$, with $\tau \approx$ 2.9 ps and $\beta \approx$ 0.67 at $\lambda$ = 0.27. For $\lambda$ approaching 1, i.e.\ the cubic structure, we can instead fit to a single exponential $e^{-(t/\tau)}$. For the cubic phase we obtain a relaxation time of $\tau \approx$ 0.69 ps.

\begin{figure}[ht!]
    \includegraphics[trim={0cm 0.1cm 0cm 0cm},clip,width=\columnwidth]{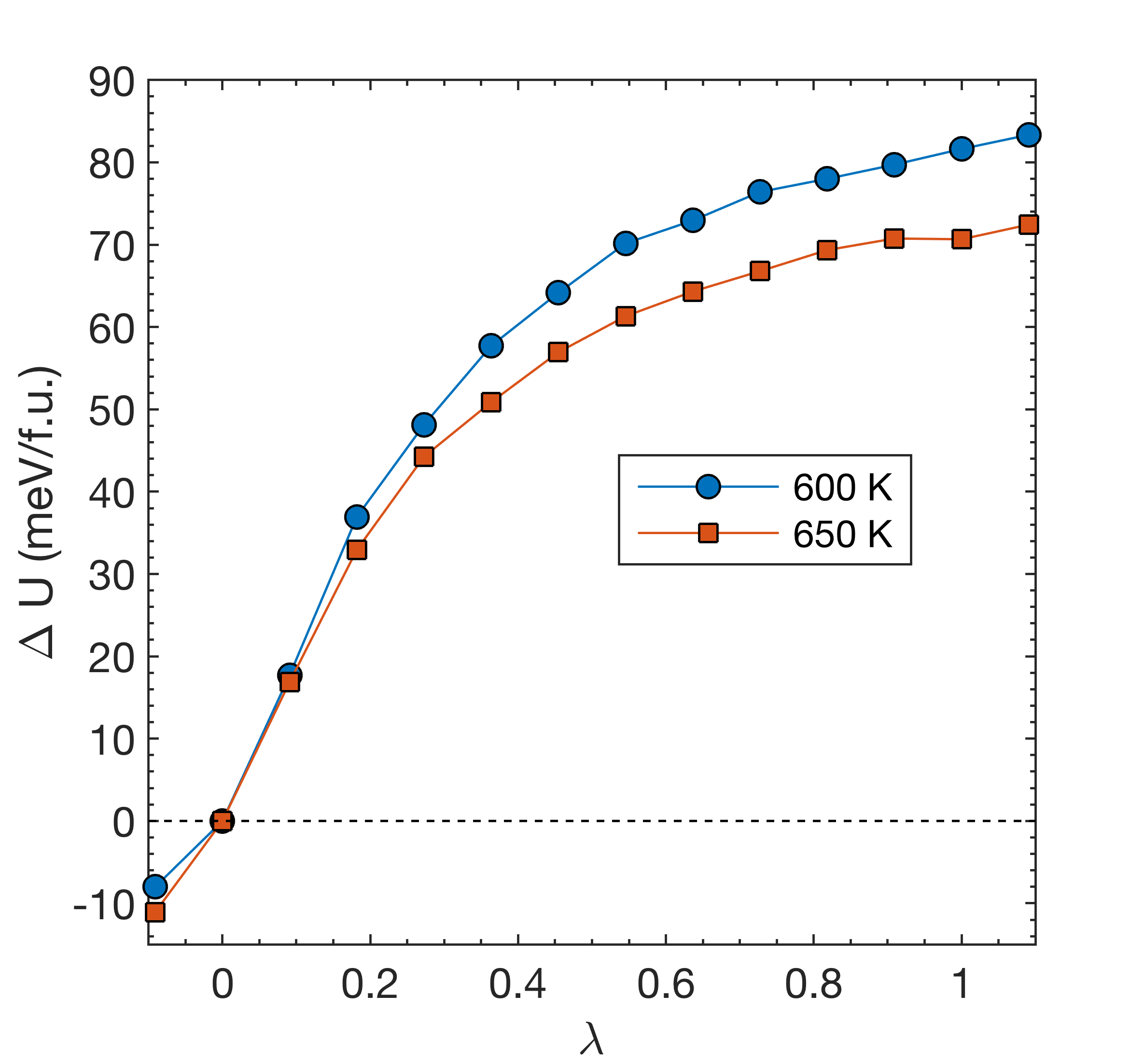}
    \caption{\label{fig:U_path} Internal energy change $\Delta U$, along the deformation path from the ordered orthorhombic phase at $\lambda$ = 0 to the dynamically disordered cubic phase at $\lambda$ = 1. Results at temperatures 600 and 650 K are shown with blue circles and red squares, respectively.}
\end{figure}

Fig.\ \ref{fig:U_path} shows the changes in internal energy $U$ along the deformation paths at 600 and 650 K. We see that the difference in $U$ between the two phases amounts to $\sim$ 82 and 71 meV/f.u.\ at 600 and 650 K, respectively. At the phase-transformation temperature this gives, by linear interpolation, a transition enthalpy of $\sim$ 80 meV/f.u.\ It is thus clear that the cubic phase is stabilised by a large entropy contribution originating from the disordered C$_2$ dimers. It is interesting to note that $U$ is increasing for $\lambda$ across 0, at 600 K the free energy minima at $\lambda$ = 0 is at least 8 meV/f.u.\ higher in $U$ compared to $\lambda \approx$ -0.09, which implies that there is a quite significant entropic part responsible for producing the free energy minima even in the orthorhombic case. Moving towards lower values of $\lambda$ corresponds to elongation of the orthorhombic $b$ axis and contraction of $a$ and $c$, which will act to suppress librational motion of the C$_2$ dimers, i.e., to increase their directional order. It thus follows that the position of cell-configuration that corresponds to the orthorhombic free energy minimum is a competition between the energetic and entropic tendencies to suppress and promote C$_2$ librations, respectively. 

Since we at this point have access to both the free and internal energy differences between the ordered and disorder phases we can extract their entropy difference $\Delta S$. 
At 600 K and 650 K, T$\Delta S$ corresponds to $\sim$ 80 and 75 meV/f.u., respectively, i.e $\Delta S$ corresponds to 0.1333 and 0.1153 meV/K/f.u., respectively.

Thus both the energy and the entropy differences between the two phases are larger at the lower temperature. This may look counter-intuitive, and lead to conceptualization that the transition is not happening due to a rapid increase of the entropy with temperature, but instead happens primarily due to an increase of the energy of the orthorhombic phase.
Let us show this would be an incomplete statement, since the changes of energy and entropy with temperature are not independent. 

If we consider the system at constant pressure and neglect the volume change with temperature, which is a reasonable approximation when comparing the states at 600 and 650 K, the thermodynamic equality following directly from a fundamental equation of thermodynamics reads
\begin{equation}
    \label{eq:UandSofTrelation}
    \left (\frac{\partial U}{\partial T}\right )_{V} = T \left (\frac{\partial S}{\partial T}\right )_{V}.
\end{equation}
Therefore, for a small temperature change from T$_{1}$ to T$_{2}$ we can write in finite differences
\begin{equation}
    \label{eq:UdiffvsSdiff}
    U(T_{2}) - U(T_{1}) = T_{1} [S(T_{2}) - S(T_{1})].
\end{equation}
Eq.\ref{eq:UdiffvsSdiff} can be easily re-written for the Helmholtz free energy as
\begin{equation}
    \label{eq:Fdiff}
    F(T_{2}) = F(T_{1}) - (T_{2} - T_{1}) S(T_{2}).
\end{equation}
If we define a difference between thermodynamic functions of the two phases, dynamically ordered (I) and disordered (II), as
\begin{equation}
    \label{eq:Diff}
\Delta X^{II-I}(T_{i}) = X^{II}(T_{i}) - X^{I}(T_{i}),
\end{equation}
where X can be F, U, or S, and i stands for either 1 or 2, then the equation for the change in the Helmholtz free energy difference between the two phases upon changing temperature from T$_{1}$ to T$_{2}$ is given by
\begin{equation}
    \label{eq:FDiff}
    \Delta F^{II-I}(T_{2}) = \Delta F^{II-I}(T_{1}) - (T_{2} - T_{1}) \Delta S^{II-I}(T_{2}).
\end{equation}
Thus, if the free energy difference between the dynamically ordered and disordered phases is positive (as we have at 600 K, i.e. $\Delta F^{II-I}(T_{1}) > 0$), then the only necessary condition for the transition into the dynamically disordered phase with increasing temperature is the positive $\Delta S^{II-I}(T_{2})$, and the size of this $\Delta S^{II-I}(T_{2})$ defines how high should be the transition temperature. We notice that such a simple estimation is very accurate in the considered case. For T$_{1}$ = 600 and T$_{2}$ = 650 K, the right-hand side of Eq. \ref{eq:FDiff} gives -3.8 meV, while its left-hand side calculated directly from TI gives -4 meV.


It is also interesting to compare the entropy between the two phases with a very simple model where the entropy difference is taken to be purely associated with a set of static configurations of the $C_2$ dimers. In the high temperature limit, we then have $\Delta S_{conf} = k_{B}\log(6)$, which gives a contribution T$\Delta S_{conf} \approx$ 93 meV/f.u.\ at T = 600 K. This is larger, but comparable, to the actual entropy contribution to the free energy difference between the two phases at 600 K of $\sim$ 80 meV/f.u. 

The C$_2$ dimers clearly have a preference of aligning along the [110] directions of the cubic cell, as seen in Fig.\ \ref{fig:struct}. However, it can also be seen from the figure that the density of dimer-alignments on paths between different [110] directions is clearly non-negligible. Furthermore, the self-relaxation time of the dimers at 600 K was estimated above to be 0.69 ps, which is comparable to the periods of atomic vibration in the system. These considerations indicate that the conceptual partitioning of the total entropy into vibrational and configurational parts, $S_{tot} = S_{vib.} + S_{conf.}$, as is commonly done for solids, is questionable in Li$_2$C$_2$, and likely other similar systems. Thus estimating, as one could typically do, the free energy by averaging the energy and harmonic vibrational entropy over many stable configurations of C$_2$ dimers and then adding an estimate of $S_{conf}$ on top of this, is not advisable for these types of systems. Instead, the free energy (or entropy) should preferably be calculated by a method which does not need to make this type of entropy partitioning, such as the stress-strain TI applied in this work.

\section{Conclusions}
To conclude we have studied the phase transformation from the low-temperature ordered orthorhombic phase to the high-temperature, dynamically disordered, cubic phase of Li$_2$C$_2$ using the stress-strain thermodynamic integration (TI) methodology. The stabilizing entropy of the cubic phase, associated with the rotational disorder of the C$_2$ dimers, is captured through the behaviour of the stress on the deformation path that connects the two phases. Using MLFF we could show that the obtained results hold in the limit of simulations times significantly longer than those accessible by AIMD. We note good agreement with available experimental results, in particular with respect to the finding that both phases appear to be local free-energy minima above as well as below the phase transition temperature. Our findings indicate that the stress-strain TI methodology can accurately describe phase transformations from ordered phases to phases with dynamical disorder related to molecular units. It may also benefit from the usage of ML techniques that allow for simulations at much larger scale.

\section{\label{sec:Acknowledgements} Acknowledgements}

The support from the Swedish Research Council (VR) (Project No. 2019-05551) and the Swedish Government Strategic Research Area in Materials Science on Advanced Functional Materials at at Link{\"o}ping University (Faculty Grant SFO-Mat-LiU No.\ 2009-00971) is acknowledged. S.I.S. acknowledges the support from the Knut and Alice Wallenberg Foundation and the ERC (synergy grant FASTCORR project 854843).  S.F. acknowledges the financial support from Carl Tryggers Stiftelse (CTS) f\"or Vetenskaplig Forskning through Grant 16:198. The computations were performed at the PDC Centre for High Performance Computing (PDC-HPC), the High Performance Computing Centre North (HPC2N) and the National Supercomputer Center (NSC),
enabled by resources provided by the Swedish National
Infrastructure for Computing (SNIC), partially funded
by the Swedish Research Council through grant agreement no. 2018-05973.

\appendix
\section{Supercell matrices}
The supercell matrices used for the orthorhombic and cubic phases are as follows. All in units of \AA.
\begin{itemize}
    \item At 600 K:
\end{itemize}
\begin{equation}
    \mathbf{h}(0) = 
\begin{pmatrix}
11.865  & 0 & 0\\
2.961& 12.229 & 0 \\
0 & 0 & 11.128
\end{pmatrix},
\end{equation}
\begin{equation}
    \mathbf{h}(1) = 
\begin{pmatrix}
11.810  & 0 & 0\\
0& 11.810 & 0 \\
0 & 0 & 11.810
\end{pmatrix}.
\end{equation}
Which corresponds to changing the lattice parameters of the conventional orthorhombic cell as:
\begin{equation}
\begin{pmatrix}
a \\
b\\
c
\end{pmatrix}
=
\begin{pmatrix}
3.764 \\
4.819\\
5.564
\end{pmatrix}
\rightarrow
\begin{pmatrix}
4.175\\
4.175\\
\sqrt2\times4.175
\end{pmatrix}
\end{equation}

\begin{itemize}
    \item At 650 K:
\end{itemize}
\begin{equation}
    \mathbf{h}(0) = 
\begin{pmatrix}
11.891  & 0 & 0\\
2.867& 12.231 & 0 \\
0 & 0 & 11.194
\end{pmatrix},
\end{equation}
\begin{equation}
    \mathbf{h}(1) = 
\begin{pmatrix}
11.82  & 0 & 0\\
0& 11.82 & 0 \\
0 & 0 & 11.82
\end{pmatrix}.
\end{equation}
\\
\\
Which corresponds to changing the lattice parameters of the convetional orthorhombic cell as:
\begin{equation}
\begin{pmatrix}
a \\
b\\
c
\end{pmatrix}
=
\begin{pmatrix}
3.784 \\
4.805\\
5.597
\end{pmatrix}
\rightarrow
\begin{pmatrix}
4.179\\
4.179\\
\sqrt{2}\times4.179
\end{pmatrix}
\end{equation}

\bibliographystyle{apsrev4-1}
\bibliography{refs.bib} 
\end{document}